\documentclass[letterpaper, 10 pt, conference]{ieeeconf}  

\IEEEoverridecommandlockouts                              

\usepackage[fleqn]{amsmath} 
\usepackage{amssymb}  
\usepackage{pgffor}
\usepackage{graphicx}
\usepackage{tikz}
\usepackage{accents}
\usepackage{siunitx}

\usetikzlibrary{%
    arrows,%
    calc,
    shapes,
    arrows,
    shapes.misc,
    shapes.arrows,%
    chains,%
    matrix,%
    positioning,
    scopes,%
    decorations.pathmorphing,
    shadows,
    arrows.meta,
    patterns
}

\newcommand\lfs{\Huge}
\newcommand\sfs{\LARGE}
\newcommand\nfs{\huge}
\newcommand\linespace{\baselineskip=20pt}
\newcommand\decisiondist{1.8cm}
\newcommand\decisiondistsmall{1.4cm}
\newcommand*\circled[1]{\tikz[baseline=(char.base)]{
    \node[shape=circle,fill=white,text=black,draw,inner sep=0.8pt] (char) {#1};}}

\newcommand\copyrighttext{%
    \footnotesize \copyright 2024 IEEE. Personal use of this material is permitted. Permission from IEEE must be obtained for all other uses, in any current or future media, including reprinting/republishing this material for advertising or promotional purposes, creating new collective works, for resale or redistribution to servers or lists, or reuse of any copyrighted component of this work in other works.}
\newcommand\copyrightnotice{%
    \begin{tikzpicture}[remember picture,overlay]
        \node[anchor=south,yshift=10pt] at (current page.south) {\fbox{\parbox{\dimexpr\textwidth-\fboxsep-\fboxrule\relax}{\copyrighttext}}};
    \end{tikzpicture}%
}

\usepackage{algorithm}
\usepackage[noend]{algpseudocode}
\let\labelindent\relax
\usepackage{enumitem}
\usepackage{booktabs}
\usepackage{cite}
\usepackage{float}
\usepackage[caption=false]{subfig}
\usepackage{mathtools}
\usepackage{array}
\usepackage[scaled=0.86]{helvet}
\usepackage{nicefrac}
\usepackage[utf8]{inputenc}
\usepackage[T1]{fontenc}
\usepackage{relsize}
\usepackage{varwidth}

\newcommand\Set[2]{\{\,#1\mid#2\,\}}
\newcommand{\vectorhoriz}[1]{
  \begin{matrix}[\,#1\,]\end{matrix}^{\top}%
}

\newcommand{\rad}{\text{rad}}

\newcommand{\acommm}{\aminforcee_{\text{comm}}}

\newcommand{\acommnew}[1]{\aminforcee_{\text{comm,new}}}

\newcommand{\aminconf}{\of{a_{\text{conf}}}{i}}

\newcommand{\timelabelleader}[2]{t^{(#1)}_{#2} }
\newcommand{\timelabel}[1]{t^{(#1)} }

\newtheorem{specification}{Specification}
\newtheorem{proposition}{Proposition}
\newtheorem{theorem}{Theorem}
\newtheorem{lemma}{Lemma}

\newtheorem{remark}{Remark}

\numberwithin{definition}{section}
\numberwithin{specification}{section}

\def\tab#1 {
    \foreach \index in {1, ..., #1} {
        \hspace{0.2cm}
    }
}
\newcommand\vtab{\vspace{0.04cm}}
\newcommand{\X}{\mathcal{X}}

\newcommand{\SRestr}{\mathcal{S}}

\newcommand{\lowerpos}[1]{\lb{\small\Call{Pos}{}}^{(#1)}}

\newcommand{\verify}{{\small\Call{Verify}{}}}
\newcommand{\upperpos}[1]{\ub{\small\Call{Pos}{}}^{(#1)}}

\algnewcommand{\LineComment}[1]{\State \(\triangleright\) #1}

\newcommand{\aleadstored}{\aminforce{\text{prec} }}
\newcommand{\aconff}{\aminforcee_{\text{conf}}}
\newcommand{\varat}[2]{#1\langle #2 \rangle}
\newcommand{\tplanning}{\Delta t_{\text{p}}}
\newcommand{\fun}[1]{\text{\textsf{#1}}}

\newcommand{\pred}[1]{\text{\small \texttt{#1}}}

\newcommand{\emgbrake}{\pred{emg}}

\newcommand{\predsafestate}{\pred{safe}}
\newcommand{\predstopbehind}{\pred{stop-before}}
\newcommand{\predcoupled}{\pred{coupled}}
\newcommand{\funcollpos}{\fun{coll}}

\newcommand{\scoll}{s_{\text{coll}}}

\newcommand{\vehi}{i}
\newcommand{\vehj}{j}

\newcommand\leader{\mathrm{l}}

\newcommand{\of}[2]{#1^{(#2)}}

\newcommand{\vwind}{v_{\text{wind}}}

\newcommand{\sensorrange}{s_{\text{sensor}}}

\newcommand{\incline}{a_{\text{incline}}}

\newcommand{\amin}[1]{\of{a_{\text{min}}}{#1}}
\newcommand{\aminphys}{a_{\text{phys}}}
\newcommand{\aminn}{a_{\text{min}}}

\newcommand{\acutin}{a_{\text{cut-in}}}

\newcommand{\amaxx}{a_{\text{max}}}

\newcommand{\aminglobal}{a_{\text{min,glob}}}
\newcommand{\amaxglobal}{a_{\text{max,glob}}}

\newcommand{\aminforcee}{\underline{a}}
\newcommand{\aminforce}[1]{\of{\aminforcee}{#1}}
\newcommand{\aminforcesub}[2]{\of{\aminforcee_{\text{#2}}}{#1}}
\newcommand{\aminforcenew}{\aminforcee_{\text{new}}}
\newcommand{\amaxforcee}{\overline{a}}

\newcommand{\vmaxx}{v_{\max}}

\newcommand{\wminn}{\underline{w}}
\newcommand{\wmaxx}{\overline{w}}
\newcommand{\wmingeneral}{\underline{w}}

\newcommand{\wmaxgeneral}{\overline{w}}

\newcommand{\adrag}{a_{\text{drag}}}

\newcommand{\aminofi}{\amin{\vehi}}

\newcommand{\sof}[1]{\of{s}{#1}}
\newcommand{\vof}[1]{\of{v}{#1}}

\newcommand{\solution}[1]{\of{\boldsymbol{\xi}}{#1}}
\newcommand{\solutionn}{\boldsymbol{\xi}}

\newcommand{\sofi}{\sof{\vehi}}

\newcommand{\solutionofi}{\solution{\vehi}}

\newcommand{\wtrajof}[1]{\of{w}{#1}}
\newcommand{\atrajof}[1]{\of{a}{#1}}
\newcommand{\atrajofi}{\atrajof{i}}

\newcommand{\wtrajofi}{\wtrajof{i}}

\newcommand{\tstep}{\Delta t_{\text{step}}}

\newcommand{\tclearremm}{\Delta t_C }

\newcommand{\lb}[1]{\underline{#1}}
\newcommand{\ub}[1]{\overline{#1}}
\algrenewcommand\algorithmicdo{}
\algrenewcommand\algorithmicthen{}

\newlength{\myeqskip}  

\let\oldcomment\Comment
\renewcommand{\Comment}[1]{\oldcomment{{\footnotesize #1}}}

\title{\LARGE \bf
Provably Correct Safety Protocol for Cooperative Platooning
}

\author{Sebastian Mair and Matthias Althoff
\thanks{All authors are with the School of Computation, Information and Technology, Technical University of
Munich, 85748 Garching, Germany}%
\thanks{{\tt\small sebi.mair@tum.de, althoff@tum.de}}%
}

\begin{document}
\bstctlcite{IEEEexample:BSTcontrol}

\setlength\abovedisplayskip{3pt}%
\setlength\belowdisplayskip{3pt}%
\setlength\abovedisplayshortskip{2pt}%
\setlength\belowdisplayshortskip{\myeqskip}
\renewcommand{\arraystretch}{0.8}

\maketitle
\thispagestyle{empty}
\pagestyle{empty}

\addtolength{\textfloatsep}{-0.3in}

\begin{abstract}

Cooperative platooning is a promising method for improving energy efficiency and traffic throughput on interstates.
Ensuring collision avoidance is particularly difficult in platooning due to the small desired inter-vehicle spacing.
We propose a safety protocol that can be applied to arbitrary controllers in platooning to prevent collisions in a
provably correct manner while still realizing a small distance to the preceding vehicle.
Our protocol intervenes as rarely and smoothly as possible, and its safety is ensured even if communication fails.
In addition, we propose a safety protocol for consensus techniques where the vehicles of the platoon successively
agree on a common braking limit.
Our safety protocols are evaluated on various scenarios using the
CommonRoad benchmark suite.

\end{abstract}
\copyrightnotice
\section{INTRODUCTION}
Platooning denotes the contactless formation of vehicles driving in a lane -- typically performed on interstates.
It aims to 1) reduce inter-vehicle distances
in order to improve energy efficiency due to diminished aerodynamic drag,
and 2) increase traffic throughput~\cite{axelsson_safety_2017}.
Platooning is often realized via cooperative adaptive cruise control (CACC), which denotes longitudinal
vehicle control incorporating information communicated between vehicles.
Due to the individual limitations of different CACC concepts~\cite{dey_review_2016}, our safety concept is
agnostic of the underlying CACC.
Moreover, it intervenes with the CACC as little as possible in order to maintain its benefits, such
as ensuring string stability.

To realize a particularly dense spacing and ensure string stability,
recent work proposes consensus techniques to establish common dynamics among heterogeneous platoons
\cite{tao_adaptive_2019, baldi_establishing_2021, liu_resilience_nodate, thormann_safe_2022}.
In particular, a consensus on the allowed acceleration interval helps to avoid large safe distances.
This is especially attractive for trucks, where the mass can significantly differ depending on the load,
affecting the individual braking capabilities.
Our safety concept includes a protocol for consensus techniques to ensure safety 1) before a consensus braking limit
is reached, and 2) when the consensus is adjusted in case the composition of the platoon changes.

\subsection{Related Work}
Safety in platooning is usually defined as the strict absence of collisions, but also a weaker safety concept based on
the potential damage caused by collisions has been proposed~\cite{alvarez_safe_1999}.
To ensure safe platooning, mainly two types of approaches are followed~\cite{mehdipour_formal_2023}:
correct-by-construction controllers and online verification.

\subsubsection{Correct-by-construction Controllers}
One group of methods designs controllers for keeping the state of platoons in invariably safe
sets~\cite{alvarez_safe_1999, scheuer_safe_2009, khalifa_vehicles_2019}.
The notion of safety can also include the avoidance of collision propagations between different
platoons~\cite{alvarez_safe_1999}.
Modelling a two-vehicle platoon as a pursuit-evasion game makes it possible to compute
the largest possible set of initial states of the following vehicle,
for which a safe controller exists~\cite{alam_guaranteeing_2014}.

\subsubsection{Online Verification}
By keeping a continuously updated fail-safe trajectory available~\cite{althoff_online_2014}, one can verify autonomous
vehicles online.
In~\cite{magdici_adaptive_2017}, an online verification concept for adaptive cruise control is presented and
evaluated for platooning, which uses fixed braking profiles as fail-safe trajectories.
The work in~\cite{ligthart_controller_2018} applies a similar approach to CACC,
and~\cite{althoff_provably-correct_2021} expands the idea in~\cite{magdici_adaptive_2017} to customizable
fail-safe trajectories and considers cut-in vehicles by solving an optimization problem online.
The work in \cite{liu_resilience_nodate} applies~\cite{althoff_provably-correct_2021} to CACC by using less
comfortable braking profiles in favor of smaller safe distances; however, this possibly leads to strong
safety interventions.
In~\cite{thormann_safe_2022}, a CACC approach based on model predictive control is extended by safety constraints.
However, the concept is not provably collision-free in continuous time.
The work also proposes that platoon vehicles communicate their braking limits to the succeeding vehicles to minimize
inter-vehicle distances. 

Apart from safety mechanisms in normal operation, dangerous situations can be defused
incorporating communication between vehicles~\cite{willke_survey_2009}.
A vehicle encountering an abnormal situation, e.g., a mechanical defect, traffic incidents, or hazardous
surface conditions,
can warn the other vehicles so that they can react accordingly~\cite{xue_yang_vehicle--vehicle_2004, reichardt_cartalk_2003}.

\subsection{Contributions}
We propose a generic safety protocol for cooperative platooning based on online verification.
In particular, we contribute the following novelties:
\begin{itemize}[leftmargin=*]
\item We prevent causing collisions in a provably correct manner under changing road inclines;
\item we enable both rare and soft safety interventions by combining strong fail-safe
maneuvers with a fallback controller;
\item we guarantee safety for braking limit changes in consensus techniques.
\end{itemize}
The rest of the paper is organized as follows:
After describing the considered system in Sec.~\ref{sec:preliminaries}, we provide an in-depth safety
specification for platooning in Sec.~\ref{sec:safety-spec}, which allows us to formulate our problem statement.
Sec.~\ref{sec:solution-concept} presents our overall safety protocol, and
Sec.~\ref{sec:consensus} describes the safety protocol for consensus techniques.
We evaluate our concept in Sec.~\ref{sec:eva}, and conclude our work
with a discussion in Sec.~\ref{sec:concl}.
%
\section{PRELIMINARIES}\label{sec:preliminaries}
%
\subsection{System Dynamics}
We address platooning on interstates with unidirectional driving and bounded road curvature~\cite{autobahn2008}.
Since the vehicles follow lanes with restricted curvature, we only consider longitudinal dynamics.
To reflect the vehicle ordering within a specific lane, we number the vehicles in ascending order, i.e., vehicle $i$
follows vehicle $i+1$.
We write $\of{\square}{i}$ to refer to the variable $\square$ of vehicle $i$, and
denote a vehicle state at time $t$ by $x(t)=\vectorhoriz{s(t) & v(t)}$, where $s$ is the front position along the road
and $v \in [0,\vmaxx]$ the velocity, with $\vmaxx$ being either the physical or legal maximum velocity.
The desired acceleration $a_d$ is the input of each vehicle, and we use $a$ for the actual acceleration.
We denote the mass of a vehicle as $m$, the drag coefficient as $c$, the frontal area as $A$, and the length as $l$.
Let us also introduce the air density $\rho$, the headwind velocity $\vwind$, and the gravity $g$.
The road incline angle at position $s$ along the road is $\alpha(s) \in [\underline{\alpha}, \overline{\alpha}]$,
where $\alpha(s) > 0$ represents an ascent.
Combining the braking capability $\aminforcee$ due to tires and brakes,
the drag $\adrag(v)=-\frac{1}{2m}\rho c A (v + \vwind)^2$,
and the acceleration $\incline(\alpha)=-g \sin(\alpha)$ caused by an incline,
the physical braking limit is~\cite[Sec. II.A]{althoff_provably-correct_2021}
\begin{equation}\label{eq:alimitsphysical}
\begin{aligned}
& \aminphys(\alpha,v)=\aminforcee + \incline(\alpha) + \adrag(v) \text{.}
\end{aligned}
\end{equation}
%
The clearing time $\tclearremm$ specifies the time within which a vehicle needs to recapture the safe distance after
a cut-in by another vehicle~\cite{maierhofer_formalization_2020, althoff_provably-correct_2021}.
\newcommand{\predcutin}[1]{\pred{cut-in}(#1)}
We use the predicate $\predcutin{i}$ to indicate that vehicle $i$ performed a cut-in within the last $\tclearremm$.
During $\tclearremm$, we assume that the cut-in vehicle does not brake harder than $\acutin$, where
this assumption is adjusted as soon as the cut-in vehicle brakes harder.
With that, the overall deceleration limit of vehicle $i$ is
\begin{equation}\label{eq:overalllimits}
\begin{aligned}
& \of{\aminn}{i}(\alpha,v) =\begin{cases}
                      \max(\acutin, \aminphys(\alpha,v)) & ~\text{if}~\predcutin{i} \\
                      \aminphys(\alpha,v) & ~\text{otherwise}\text{.}
                 \end{cases} \\
\end{aligned}
\end{equation}
%
We introduce the disturbance $w \in [\wminn,\wmaxx]$, where the disturbance set $[\wminn,\wmaxx]$ can be used to
compensate model inaccuracies using reachset conformance~\cite{schurmann_ensuring_2017}.
Given the maximum possible acceleration $\amaxx(s, v)$ due to engine characteristics, drag, and incline,
the vehicle dynamics can be written as~\cite[Eq. (1)]{althoff_provably-correct_2021}
\begin{equation}
\begin{aligned}
& \dot{s} = v \\
& \dot{v} = \begin{cases}
                                  0 & \text{if}~(v \leq 0 \wedge a_d + w \leq 0) \\
                                    & \vee (v \geq \vmaxx \wedge a_d + w \geq 0)  \\
                                  \aminn(\alpha(s), v) + w & \text{if}~a_d < \aminn(\alpha(s), v) \\
                                  \amaxx(\alpha(s), v) + w & \text{if}~a_d > \amaxx(\alpha(s), v) \\
                                  a_d + w & \text{otherwise} \text{.}
                               \end{cases}
\end{aligned}
\label{eq:model}
\end{equation}
%
For an initial state $x_0$, an input trajectory $a_d(\cdot)$, and a disturbance trajectory $w(\cdot)$, the solution
of the model in~\eqref{eq:model} over time $t$ is denoted as $\solutionn(t;x_0,a_d(\cdot),w(\cdot))$.
We write $\solutionn_s$ to refer to the position of $\solutionn$.
We allow the input $a_d=-\infty$, which results in full braking according to~\eqref{eq:model}.
%

\subsection{Platoon Vehicle Assumptions}
In addition to measuring its state variables, each vehicle in the platoon obtains the relative position and velocity of
each preceding vehicle within a sensor range of at least $\sensorrange$.
To account for measurement uncertainties, we assume that a measurement results in an interval
$[\underline{\square},\overline{\square}]$ enclosing the actual value of
variable $\square$~\cite{althoff_provably-correct_2021}.
Measurement intervals are also provided to each platoon vehicle for $\alpha(s)$, $\rho$, and $\vwind$.
%

Furthermore, we assume that each platoon vehicle $i$ is equipped with an arbitrary CACC,
denoted as the \textit{nominal} CACC of vehicle $i$.
It operates with a planning period of $\of{\tplanning}{i}$, i.e., it provides a new input $a_d$
at discrete planning times $t_k=k\of{\tplanning}{i}$ to be applied in $[t_k, t_{k+1})$.
\section{SAFETY SPECIFICATION AND PROBLEM STATEMENT}\label{sec:safety-spec}
In this section, we introduce a comprehensive notion of safety in platooning,
and subsequently formulate our problem statement.

\subsection{Safety Specification}
First, we introduce relevant predicates and a function referring to vehicle $i$ at time $t$.
For brevity, we omit the dependence on $i$ and $t$ in the notation.
\begin{itemize}
      \item $\emgbrake$: Vehicle $i$ brakes as strong as possible at time $t$.
      To prevent potential collisions within the platoon, the vehicle communicates a possible imminent collision
      backward within the time interval $[t,t+\of{\tplanning}{i}]$,
      containing its predicted rear position at the time of the collision.
\item $\predsafestate(j)$: Vehicle $i$ is safe w.r.t. vehicle $j > i$ at time $t$, formally defined by
      \begin{equation}
      \begin{aligned}\label{eq:safe-def}
      & \predsafestate(j) \iff  \exists \atrajofi(\cdot)\,
      \forall \wtrajofi(\cdot)\, \forall \atrajof{j}(\cdot)\, \forall \wtrajof{j}(\cdot)\,
      \forall t' \geq 0\colon \\
      & \tab{5} \solutionofi_s(t'; \of{x}{i}(t),\atrajofi(\cdot),\wtrajofi(\cdot)) \\
      & \tab{5}   < \solution{\vehj}_s(t';\of{x}{j}(t),\atrajof{\vehj}(\cdot),\wtrajof{\vehj}(\cdot)) - \of{l}{j}
      \text{.}
      \end{aligned}
      \end{equation}
\item $\predstopbehind(s)$: Vehicle $i$ can stop before position $s$.
For a standing vehicle $j$ with rear position $s$, the predicate is defined by
$\predstopbehind(s) \iff \predsafestate(j)$. 
\item $\funcollpos(j)$: The function returns the collision position that vehicle $i$ received by vehicle $j$
as part of a collision alert before time $t$.
If vehicle $i$ did not receive a collision alert or if it was withdrawn by vehicle $j$, $\infty$ is returned.
\end{itemize}
%

Using first-order logic, we now provide formal specifications to guarantee legal safety in
platooning~\cite{althoff_online_2014}.
We use the symbol $\veebar$ to denote the exclusive disjunction operation.
\begin{specification}[Stopping within Sensor Range~\cite{althoff_provably-correct_2021}]\label{spec:stop-in-fov}
Vehicle $i$ is always able to stop within its sensor range:
\begin{equation}
\begin{aligned}
\forall t \colon \predstopbehind(\sofi(t) + \sensorrange) \text{.}
\end{aligned}
\end{equation}
\end{specification}

\begin{specification}[Collision Avoidance]\label{spec:safety}
Vehicle $i$ has to keep a safe distance unless another vehicle performed a cut-in,
triggering the emergency procedure:
\begin{equation}
\begin{aligned}
& \forall t\, \forall j > i\colon \predsafestate(j) \veebar \emgbrake
\text{.}
\end{aligned}
\end{equation}
\end{specification}
%

\begin{specification}[Multiple Collision Avoidance]\label{spec:coll-pileup}
When receiving a collision alert, vehicle $i$ must stop before the collision or trigger the emergency procedure:
\begin{equation}
\begin{aligned}
& \forall t\, \forall j > i\colon \predstopbehind(\funcollpos(j)) \veebar \emgbrake \text{.}
\end{aligned}
\end{equation}
\end{specification}
\smallskip
Based on these safety specifications, we formulate the problem statement next.

\subsection{Problem Statement}
The objective of this work is to develop a safety protocol applicable to CACC and consensus techniques in platooning,
which ensures that each vehicle of the platoon always fulfills the
specifications~\ref{spec:stop-in-fov}-\ref{spec:coll-pileup}, while
1) still realizing small inter-vehicle distances,
2) intervening with the nominal CACC as rarely and smoothly as possible,
3) handling communication failures, and
4) ensuring convergence of the consensus techniques.
\section{SAFETY PROTOCOL}\label{sec:solution-concept}
This section describes the overall safety protocol;
the protocol for consensus techniques is explained later.
During platoon formation, adjacent vehicles $i-1$ and $i$ perform a handshake and mutually confirm that they apply the
safety protocol.
Furthermore, vehicle $i$ communicates its parameters to vehicle $i-1$.
We use the predicate $\predcoupled(i-1, i)$ to denote that vehicles $i-1$ and $i$ successfully coupled.
\begin{figure}[t]
    \centering
    \vspace*{0.3cm}
    \hspace{-1.5cm}
    \resizebox{1.15\columnwidth}{!}{\input{safety-protocol-overview.tikz}}
    \vspace{-0.2cm}
    \caption{Overview of our safety protocol without consensus for vehicle $i$, including communication (dashed).}
    \label{fig:overview-safety-algo}
    \vspace{0.5mm}
\end{figure}

Fig.~\ref{fig:overview-safety-algo} provides an overview of the protocol:
Every vehicle $i$ verifies compliance with the specifications~\ref{spec:stop-in-fov}-\ref{spec:coll-pileup}
in each planning step.
The planned input $a_d$ is only applied if it is compliant.
Otherwise, a fallback controller is engaged.
In some cases, vehicle $i$ must communicate an imminent collision with its succeeding vehicles.
\subsection{Solution Concept}\label{subsec:safety_verification}

\begin{figure}[t]
\centering
\vspace*{0.1cm}
\includegraphics[width=1.0\columnwidth, trim={1cm 0.5cm 0.7cm 0.5cm},clip]{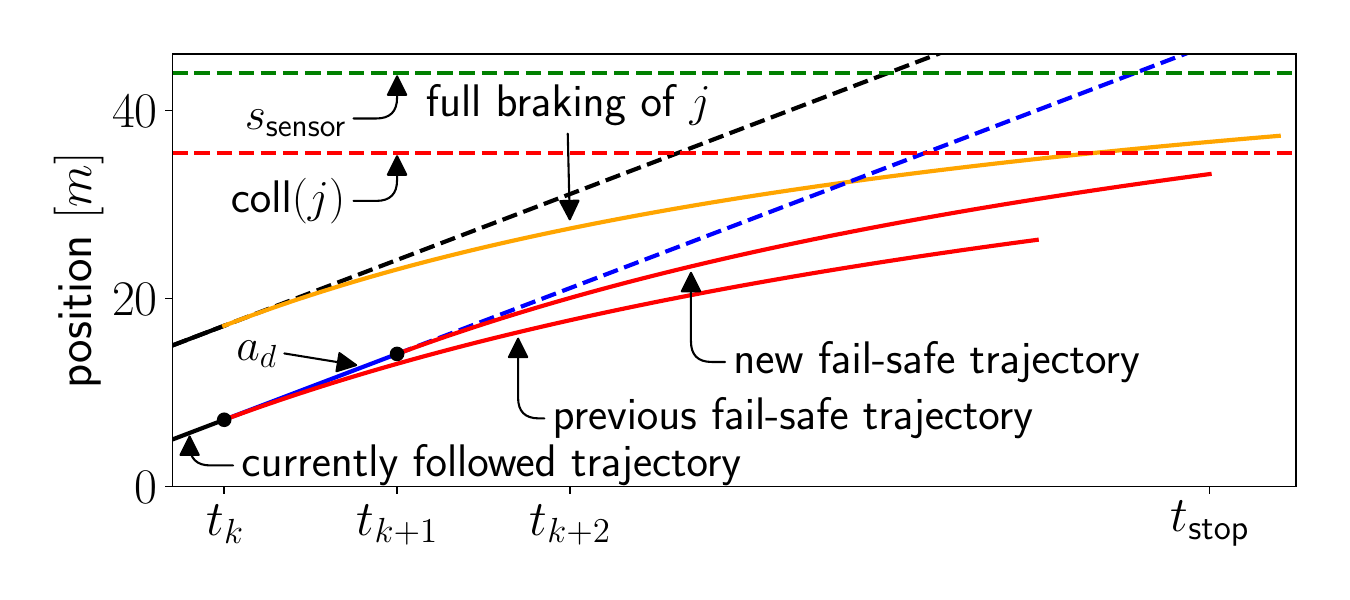}
\vspace{-0.3cm}
\caption{Vehicle $i$ successfully verifies an input $a_d$ in the $k$-th planning step w.r.t. a preceding vehicle $j$.
The dashed blue and black lines represent possible future trajectories of vehicle $i$ and $j$, respectively.}
\label{fig:fail_safe_1}
\end{figure}

\newcommand{\tstop}{t_{\text{stop}}}
We describe how a vehicle $i$ verifies if its planned input $a_d$ complies with
Spec.~\ref{spec:stop-in-fov}-\ref{spec:coll-pileup}.
Fig.~\ref{fig:fail_safe_1} sketches the approach, which extends the concept in~\cite{althoff_provably-correct_2021}
by collision alerts:
Vehicle $i$ simulates its own state forward in time for the time $\of{\tplanning}{i}$,
followed by a full brake as a fail-safe trajectory until standstill at time $\tstop$.
It also simulates each preceding vehicle $j > i$ until $\tstop$, assuming that vehicle $j$ immediately performs a full
brake.
Here, vehicle $i$ assumes worst-case parameters for vehicle $j$.
Only for vehicle $i+1$, vehicle $i$ utilizes the parameters received via communication (cf. Fig.~\ref{fig:overview-safety-algo})
to enable a smaller safe distance.
The input $a_d$ is classified as safe if vehicle $i$ always stays behind
a) its sensor range (cf. Spec.~\ref{spec:stop-in-fov}),
b) each vehicle $j$ (cf. Spec.~\ref{spec:safety}), and
c) the received collision positions (cf. Spec.~\ref{spec:coll-pileup}).
If the verification fails in the next planning cycle, vehicle $i$ can still safely apply full braking
from $t_{k+1}$ on.
This verification procedure guarantees compliance with Spec.~\ref{spec:stop-in-fov}-\ref{spec:coll-pileup},
which follows by induction over the planning steps
assuming that $\vof{i}(t_0)=0$~\cite[Sec. IV.A)]{althoff_provably-correct_2021}. 

\newcommand{\deltatsim}{\Delta t_{\text{sim}}}
Note that the simulation cannot be computed exactly due to measurement uncertainties,
unknown future disturbances, and a missing closed-form solution of $\solutionn_s$.
To still guarantee safety, we compute upper bounds $\upperpos{i}(a_d(\cdot), t)$
for the front position of vehicle $i$ for future time points $t$, and lower bounds $\lowerpos{j}(a_d(\cdot), t)$ for the
rear positions of the preceding vehicles, as detailed in Sec.~\ref{sec:monotonicity}.

Alg.~\ref{alg:safety-algo} summarizes the described verification procedure.
The time discretization $\deltatsim$ for the forward simulation can be chosen by the user (l. 4).
Note that during a time interval $[r\deltatsim, (r+1)\deltatsim]$,
we use the left limit $r\deltatsim$ for the preceding vehicles and the right limit $(r+1)\deltatsim$
for vehicle $i$ (l. 4) to guarantee safety over each consecutive time interval.
\newcommand{\scollset}{\SRestr_{\text{coll}}}
\algrenewcommand\algorithmicindent{0.9em}%
\begin{figure}[!t]
    \vspace{-4.5mm}
\begin{algorithm}[H]
\small
\caption{\textsc{Verify}$(a_d, \X)$}
\label{alg:safety-algo}
\textbf{Input:} Input $a_d$, preceding vehicles $\X=\{i+1, \dots, N\}$ \\
\textbf{Output:} Boolean indicating if vehicle $i$ can safely apply $a_d$
\begin{algorithmic}[1]
\State $a_{d,\text{ego}}(t) = \begin{cases}
        a_d & \text{if}~t\leq \of{\tplanning}{i}  \\
        -\infty & \text{otherwise}\text{.}
    \end{cases}$ \Comment{simulated traj. for vehicle $i$}
\vtab \State $a_{d,\text{prec}}(t) = -\infty$ \Comment{simulated traj. for preceding vehicles}
\State stop-in-sensor-range $\gets \upperpos{i}(a_{d,\text{ego}}(\cdot), \tstop) < \sensorrange$
\vtab \vtab \State $\begin{aligned}[t]
                  & \text{no-coll} \gets \forall j \in \X\,
                  \forall r \in \Set{r \in \mathbb{N}_0}{(r-1)\deltatsim\leq\tstop } \colon \\
                          & \upperpos{i}(a_{d,\text{ego}}(\cdot), (r+1)\deltatsim) < \lowerpos{j}(a_{d,\text{prec}}(\cdot), r\deltatsim)
\end{aligned}$

\vtab \State no-multi-coll $\gets \forall j \in \X\colon
\upperpos{i}(a_{d,\text{ego}}(\cdot), \tstop) < \funcollpos(j)$

\vtab \State \Return $\text{stop-in-sensor-range}~\wedge~\text{no-coll}~\wedge~\text{no-multi-coll}$
\end{algorithmic}
\end{algorithm}
\vspace{-1.5mm}
\end{figure}
%
\subsection{Overall Protocol}\label{subsec:overall_approach}

The overall safety protocol for a single planning step is summarized in Alg.~\ref{alg:overall-algo}.
%
\algrenewcommand\algorithmicindent{0.9em}%
\begin{figure}[!t]
    \vspace{-2mm}
\begin{algorithm}[H]
\small
\caption{\textsc{SafetyProtocol}$(\X)$}
\label{alg:overall-algo}
\textbf{Input:} Preceding vehicles $\X=\{i+1, \dots, N\}$ \\
\textbf{Output:} Input acceleration $a_d$ for vehicle $i$ in $[t_k, t_{k+1})$
\begin{algorithmic}[1]
\If{$\predcoupled(i, i+1)\colon$}$\X \gets \{i+1\}$ \Comment{vehicle elimination}
\EndIf
\State $a_{0} \gets~\Call{SafeConsensus}{\X}$ \Comment{Alg.~\ref{alg:update-a-min} in Sec.~\ref{sec:consensus}}
\State $a_{1} \gets \Call{Cut-inHandling}{i, \X}$ \Comment{cf.~\cite[Sec. V.B.2)]{althoff_provably-correct_2021}}
\State $a_{2} \gets \Call{NominalCACC}{\X}$
\State $a_d \gets~\min(\{a_0, a_1, a_2\})$
\vtab \State $\text{desired-safe} \gets \verify(a_d, \X)$ \Comment{Alg.~\ref{alg:safety-algo} in Sec.~\ref{subsec:safety_verification}}
\If{$\neg \text{desired-safe}\colon$}
\State $(a_d, \text{fallback-safe}) \gets \Call{Fallback}{i, \X}$
\vtab \If{$\neg\text{fallback-safe}\colon$} $\Call{sendAlert}{ \scoll }$
\EndIf
\EndIf
\If{$\text{desired-safe} \vee \text{fallback-safe}\colon$} $\Call{withdrawAlert}{}()$
\EndIf
\State \Return $a_d$
\end{algorithmic}
\end{algorithm}
    \vspace{-0.3cm}
    \end{figure}
If vehicles $i$ and $i+1$ are coupled, then vehicle $i$ can safely ignore all vehicles in front of vehicle $i+1$ (l. 1),
which holds by induction over the preceding platoon vehicles.
%
We then invoke the protocol for consensus techniques (l. 2, cf. Sec.~\ref{sec:consensus}).
To ensure an appropriate reaction to cut-in vehicles, we integrate an approach
similar to~\cite[Sec. V.B.2)]{althoff_provably-correct_2021} (l. 3). 
Both the $\Call{SafeConsensus}{}$ and $\Call{Cut-inHandling}{}$ subroutine return an upper bound for the acceleration.
Thus, the desired acceleration is the nominal input (l. 4) limited by these two values (l. 5).
We verify safety of the desired acceleration (l. 6) (cf. Sec.~\ref{subsec:safety_verification}).
To avoid full braking whenever possible, we additionally engage a fallback controller in case the desired acceleration
is unsafe, which computes the maximum
acceleration that is still safe using binary search (l. 8).
This is a simple yet effective approach, as it can be run anytime-like in parallel to the nominal
CACC by continuously improving the solution accuracy.
In case of a cut-in or a received collision alert, even the previous fail-safe maneuver
can become unsafe.
The fallback controller then returns $a_d=-\infty$ and $\text{fallback-safe}=\text{false}$, and
the imminent collision is communicated with the predicted position of collision
$\scoll$ (l. 9). 
Otherwise, collision alerts sent previously are withdrawn (l. 10).
\subsection{Computing Bounds on Reachable Positions}\label{sec:monotonicity}
We now present the computation of bounds on the reachable positions of a vehicle, denoted by the functions
$\lowerpos{j}(a_d(\cdot), t)$ and $\upperpos{i}(a_d(\cdot), t)$.
We utilize a specific type of monotonicity for this;
the dynamics of our vehicle model in~\eqref{eq:model} is not monotone in a classical sense~\cite{angeli_monotone_2002},
because the change in velocity is position-dependent due to the road incline.
However, we show that under the assumption~\eqref{eq:ass-a-non-incr} on $a_d(\cdot)$, monotonicity in the position
domain holds.

%
\begin{theorem}[Monotonicity in the Position Domain]
    \label{prop:monotonicity}
    For any input trajectories $\lb{a}_d(\cdot)$, $a_d(\cdot)$, and $\ub{a}_d(\cdot)$ fulfilling
    \begin{equation}
        \label{eq:ass-a-non-incr}
        \begin{aligned}
            a_d(\cdot)~\text{is non-increasing over}~t \text{, and}
        \end{aligned}
    \end{equation}
    \begin{equation}
        \label{eq:ass-a-bounds}
        \begin{aligned}
            \forall t\colon \lb{a}_d(t) \leq a_d(t) \leq \ub{a}_d(t) \text{,}
        \end{aligned}
    \end{equation}

    \vspace{0.1cm}
    it holds that
    \vspace{0.1cm}
    \begin{equation}
        \label{eq:monotonicity}
        \begin{aligned}
            \forall t \geq 0\colon & \solutionn_s(t;\underline{x},\lb{a}_d(\cdot),\wminn) \\
            & \leq \solutionn_s(t;x,a_d(\cdot),w(\cdot)) \\
            & \leq \solutionn_s(t;\overline{x},\ub{a}_d(\cdot),\wmaxx)\text{.}
        \end{aligned}
    \end{equation}

    \proof
    {
        We only show the first inequality in~\eqref{eq:monotonicity}, as the proof for the second one works analogously.
    For brevity, we write $\lb{s}(t)$ and $\lb{v}(t)$ for the position and velocity of
        $\solutionn(t;\underline{x},\lb{a}_d(\cdot),\wminn)$, and $s(t)$ and $v(t)$ analogously for
        $\solutionn(t;x,a_d(\cdot),w(\cdot))$.
    First, we show that for any $t_a \leq t_b$ with $\lb{s}(t_b)=s(t_a)$ and $\lb{v}(t_b)=v(t_a)$,
        it holds that
        \begin{equation}
            \label{eq:corollary-v-dot}
            \begin{aligned}
                \dot{\lb{v}}(t_b) \leq \dot{v}(t_a) \text{.}
            \end{aligned}
        \end{equation}
        From~\eqref{eq:ass-a-non-incr} and~\eqref{eq:ass-a-bounds} follows that $\lb{a}_d(t_b) \leq a_d(t_a)$,
        which, together with~$\wminn \leq w(t_a)$, results in~\eqref{eq:corollary-v-dot}
    according to the dynamics in~\eqref{eq:model}.

    Let us assume that the first inequality in~\eqref{eq:monotonicity} does not hold for the sake of contradiction.
    The continuity of $s$ in $t$ implies that there is a minimum $t' > 0$, 
        such that $\lb{s}(t') = s(t')$
        and $\exists \epsilon>0~\forall \delta \in (0, \epsilon]\colon \lb{s}(t' + \delta) > s(t' + \delta)$.
    Thus
        \begin{equation}
            \label{eq:implication-speed}
            \begin{aligned}
                \lb{v}(t') > v(t') \text{,}
            \end{aligned}
        \end{equation}
        as $\lb{v}(t')=v(t')$ implies that $\dot{\lb{v}}(t') > \dot{v}(t')$, contradicting~\eqref{eq:corollary-v-dot}.
    From~\eqref{eq:implication-speed},
        $\underline{x} \leq x$, 
        and the continuity of $v$ in $s$
        follows that there must be a position $s'$, such
    that there are $t_a, t_b$ with $t_a \leq t_b < t'$
        and $\lb{s}(t_b)=s(t_a)=s'$, $\lb{v}(t_b)=v(t_a)$ and
        $\exists \epsilon>0\,\forall \delta \in (0, \epsilon]\colon \lb{v}(t_b + \delta) > v(t_a + \delta)$.
    This implies that $\dot{\lb{v}}(t_b) > \dot{v}(t_a)$, which contradicts~\eqref{eq:corollary-v-dot},
        thus, the assumption on $t'$ must be wrong, proving the theorem. $\square$
    }
\end{theorem}

For a time step size $\tstep \geq 0$ and an input trajectory $a_d(\cdot)$ fulfilling~\eqref{eq:ass-a-non-incr},
we choose $\lb{a}_d(t)=a(t+\tstep)$, and $\ub{a}_d(\cdot)$ is the zero-order hold function of $a_d(\cdot)$.
Theorem~\ref{prop:monotonicity} allows us to compute
$\lowerpos{j}(a_d(\cdot),t)$ and $\upperpos{i}(a_d(\cdot), t)$
using $\lb{a}_d(\cdot)$ and $\ub{a}_d(\cdot)$, respectively, with standard solvers for ordinary differential equations (see
Appendix for further details).
%
%
\section{SAFETY PROTOCOL FOR CONSENSUS TECHNIQUES}\label{sec:consensus}
\newcommand{\acons}{\underline{a}_{\text{cons}}}
We additionally consider the possibility that all vehicles in the platoon use a consensus scheme to establish a
common braking limit $\acons$ that is not known a priori;
this value obviously changes when vehicles enter or leave the platoon.
We assume that the consensus scheme runs black-box entities on all vehicles in the platoon,
denoted as \textit{consensus entities}, that keep proposing new braking limits for their vehicle,
which gradually converge to $\acons$.
Both decentralized and centralized consensus schemes can be represented in this way, as well as
schemes that establish a consensus braking limit successively and those that only require a single change.

%
Our protocol in Fig.~\ref{fig:overview-consensus-algo} for safely changing the braking limit
using consensus techniques is embedded in
the overall safety protocol as the $\Call{SafeConsensus}{}$ procedure (cf. l. 2 in Alg.~\ref{alg:overall-algo}).
For each new braking limit $\aminforcenew$ provided by its consensus entity
(step 1 in Fig.~\ref{fig:overview-consensus-algo}),
vehicle $i$ verifies if safety is still upheld (step 2).
If the safety of the succeeding vehicle is affected by the new braking limit,
vehicle $i$ requests a safety confirmation by vehicle $i-1$ for $\aminforcenew$ (steps 3-4).
Only if safety is ensured, vehicle $i$ adopts $\aminforcenew$ as the new braking limit.
\begin{figure}[t]
\centering
\vspace*{0.3cm}
\hspace{-1.5cm}
\resizebox{1.15\columnwidth}{!}{\input{consensus-overview.tikz}}
\vspace{-0.15cm}
\caption{
    Sketch of safely updating braking limits from the perspective of vehicle $i$,
    including communication (dashed).
}
\label{fig:overview-consensus-algo}
\end{figure}

\subsection{Overview}
We present an overview of the $\Call{SafeConsensus}{}$ procedure in Fig.~\ref{fig:details-consensus-algo} using an activity
diagram:
\begin{figure}[t]
    \vspace*{0.3cm}
\centering
\resizebox{1.\columnwidth}{!}{

\tikzset{
    general/.style = {fill=black!3, draw, minimum width = 0.5em, minimum height = 0.5em, align=center},
    process/.style    = {general, rounded corners=0.5em, inner sep=0.6em},
    object/.style    = {general, rectangle, inner sep=0.6em},
    decision/.style      = {general, diamond, aspect=1.3, inner sep=0, text width=4.2em},
    successor/.style = {pattern=north west lines, pattern color=black!30},
    pin/.style    = {fill=black!5,draw, thick, rectangle, minimum height = 0.6em,
    minimum width = 0.6em, node distance=-1pt, inner sep=0,font=\relsize{-3.5}},
    start/.style      = {fill=black,draw,circle,node distance=4em,minimum size=0.4cm},
    finalInner/.style      = {fill=black,draw,circle, minimum size=0.2cm},
    final/.style      = {draw=black,thick,fill=white,circle,minimum size=0.5cm},
    group/.style      = {color=black,thin,rounded corners=0.8em, rectangle},
    groupCaption/.style      = {above=0.2cm,right=0.2cm,fill=white},
    input/.style    = {coordinate,node distance=2em},
    output/.style   = {coordinate,node distance=2em},
    helper/.style   = {outer sep=0, inner sep=0, minimum size=0cm, circle},
    arr/.style   = {-{Latex[length=3mm]}},
    comm/.style = {arr, dash pattern=on 5pt off 5pt}, 
    legendline/.style   = {line width=0.75mm},
    between/.style args={#1 and #2}{ 
        at = ($(#1)!0.5!(#2)$)
    }
}

\begin{tikzpicture}[
        node distance=1cm,
        every node/.style={inner sep=0,outer sep=0},
        label/.style={font=\sffamily},
    ]
    \newcommand{\xshift}{1.5cm}
\draw
    node at (0, 0) [start](Start){}
    node [process, right=1.5cm of Start](ConsensusEntity){\nfs Apply consensus entity of veh. $i$}
    node [decision, below=of ConsensusEntity, text width=9em](aNewLarger){\nfs $\aminforcenew<\aminforcee$}
    node [below left=3cm of aNewLarger,xshift=-1.2cm-\xshift, yshift=0.8cm](HelperCaseA){}
    node [below right=2cm of aNewLarger,xshift=3cm-\xshift, yshift=0.8cm](HelperCaseB){}
    node [left=6cm of aNewLarger, yshift=-1.5cm, xshift=-\xshift](UpperLeftCaseA){}
    node [right=0cm of aNewLarger, yshift=-1.5cm, xshift=-\xshift](UpperLeftCaseB){}
;

\draw[arr](Start) -- (ConsensusEntity);
\draw[arr](ConsensusEntity) -- (aNewLarger);
\draw
    node [decision, successor, text width=10em, below= of HelperCaseA](IsALeadSafe){\nfs \linespace Is veh. $i-1$ safe if veh. $i$ uses $\aminforcenew$?\par}
    node [process, successor, below=\decisiondist of IsALeadSafe, text width=10em, yshift=0.3cm](BrakeSofty2){\nfs \linespace Increase distance to front \par}
    node [process, below=of BrakeSofty2](OvertakeANewAfterConf){\nfs Change $\aminforcee \gets \aminforcenew$}
    node [right=1.5cm of OvertakeANewAfterConf](Helper2){}
    node [below right=of OvertakeANewAfterConf, xshift=1.7cm](BottomRightCaseA){}
    ;
\draw[comm](aNewLarger) -| (IsALeadSafe) node [pos=0.5, anchor=south, xshift=2mm] {\nfs [yes] }
    node [pos=0.8, anchor=west, xshift=-0.1cm, yshift=-0.7cm, text width=3.5cm, align = right] {\sfs Safety request \\ for $\aminforcenew$};
\draw[arr](IsALeadSafe) -- (BrakeSofty2) node [pos=0.5, anchor=east, xshift=-2mm] {\nfs [no] };
\draw[comm](IsALeadSafe) -| (Helper2) node [pos=0.5, anchor=south, yshift=2mm] {\nfs [yes] }
    node [pos=0.63, anchor=east, xshift=-0.2cm, yshift=-0.7cm, align=right] {\sfs Safety\\ \sfs confirm.\\ \sfs for $\aminforcenew$}
    -- (OvertakeANewAfterConf);

\draw [group](UpperLeftCaseA)rectangle(BottomRightCaseA);
\node at (UpperLeftCaseA -| BottomRightCaseA) [groupCaption, minimum width=3.2, xshift=0.2cm-3.5cm] {\nfs \textbf{\textsc{Case A}}};

\draw
node [decision, below= of HelperCaseB, text width=10em](IsANewSafe){\nfs \linespace Is veh. $i$ safe if
using $\aminforcenew$? \par}
node [process, below=\decisiondistsmall of IsANewSafe](OvertakeANew){\nfs Change $\aminforcee \gets \aminforcenew$}
node [process, below=of OvertakeANew, xshift=4cm, text width=7em](BrakeSofty){\nfs \linespace Increase distance to front \par}
node [above=of BrakeSofty](Helper3){}
node [below right=of BrakeSofty, xshift=-0.3cm](BottomRightCaseB){}
;
\draw[arr](aNewLarger) -| (IsANewSafe) node [pos=0.5, anchor=south, yshift=2mm] {\nfs [no] };
\draw[arr](IsANewSafe) -| (Helper3) node [pos=0.4, anchor=south, yshift=2mm] {\nfs [no] } -- (BrakeSofty);
\draw[arr](IsANewSafe) -- (OvertakeANew) node [pos=0.5, anchor=west, xshift=2mm] {\nfs [yes] };

\draw [group](UpperLeftCaseB)rectangle(BottomRightCaseB |- BottomRightCaseA);
\node at (UpperLeftCaseB) [groupCaption, minimum width=3.2, xshift=0.2cm] {\nfs \textbf{\textsc{Case B}}};

\end{tikzpicture}}
\vspace{-0.1cm}
\caption{
    Activity diagram of the protocol for consensus techniques from the perspective of
    vehicle $i$, including communication (dashed) and activities carried out by vehicle $i-1$ (hatched).
}
\label{fig:details-consensus-algo}
\vspace{1mm}
\end{figure}
The consensus entity of vehicle $i$ provides a new braking limit $\aminforcenew$ in the $k$-th planning step.
Vehicle $i$ 
distinguishes two cases:

\paragraph{Case A ($\aminforcenew < \aminforcee$)}
Safety of vehicle $i$ w.r.t. the preceding vehicles would not be changed
by $\aminforcenew$, but safety of the coupled succeeding vehicle $i-1$ due to stronger braking capabilities
of the preceding vehicle.
Therefore, vehicle $i$ requests a safety confirmation for $\aminforcenew$ from vehicle $i-1$.
Vehicle $i-1$ receives the confirmation request and re-verifies its safety,
assuming that vehicle $i$ was using $\aminforcenew$ instead of $\aminforcee$.
If safety is not provided, vehicle $i-1$ starts to increase the distance to vehicle $i$.
As soon as safety is ensured, vehicle $i-1$ sends a confirmation to vehicle $i$ so that
vehicle $i$ can safely change its currently used braking
limit to $\aminforcenew$.
\paragraph{Case B ($\aminforcenew \geq \aminforcee$)}
An increased braking limit requires vehicle $i$ to re-verify its safety w.r.t. the preceding vehicles.
As soon as the verification is successful, it can change its braking limit to $\aminforcenew$.
Otherwise, it starts to increase the distance to the front, such that
$\aminforcenew$ can be adopted at some point in the future.
By this, convergence of the consensus scheme is ensured.


%
\subsection{Protocol in Detail}
\newcommand{\tconff}{t_{\text{conf}}}
\newcommand{\alead}{\aminforcesub{\text{prec}}{\text{comm}}}
\newcommand{\tlead}{\timelabelleader{\text{prec}}{\text{comm}} }
\newcommand{\tcomm}{t_{\text{comm}} }
\newcommand{\tleadstored}{\timelabel{\text{prec}}}
Alg.~\ref{alg:update-a-min} shows the complete $\Call{SafeConsensus}{}$ protocol.
%
\newcommand{\transact}{\text{increase-dist}}
\newcommand{\getatrans}{\Call{GetInputBound}{}}
\algrenewcommand\algorithmicindent{0.9em}%
\begin{figure}[!t]
    \vspace{-2mm}
\begin{algorithm}[H]
\small
\caption{\textsc{SafeConsensus}$(\X)$}
\label{alg:update-a-min}
\textbf{Persistent variables:}
\begin{itemize}
\item $\aminforcee$: currently used braking limit of vehicle $i$, initially set to the physical limit.
\item $\aleadstored$: braking limit for vehicle $i+1$, initially chosen based on worst-case assumptions.
\item $\acommm$: latest communicated braking limit of vehicle $i$, initially set to $-\infty$.
\end{itemize}
\textbf{Input}:
Received time-labeled data:
\begin{itemize}
\item From vehicle $i-1$ (sent in l. 19): $\aconff$ (braking limit of vehicle $i$ for which
vehicle $i-1$ confirms its safety).
If no message has been received: $\aconff \gets \infty$.
\item From vehicle $i+1$ (sent in l. 12): $\alead$ (new braking limit
that vehicle $i+1$ either requests a safety confirmation for or already adopted safely).
\end{itemize}
\textbf{Output}: Upper bound $a_{\text{trans}}$ on input acceleration
\begin{algorithmic}[1]
\State $\transact \gets \text{false}$
\LineComment{\textbf{Apply consensus scheme:}}
\State $\aminforcenew \gets~\Call{ConsensusEntity}{}()$
\vtab
\If{$\aminforcenew \geq \aminforcee\colon$} \Comment{Case B}
\vtab   \If{$\verify(-\infty,\X)$ using $\aminforcenew$ for $\aminforcee\colon$}
            \State $\aminforcee \gets \aminforcenew$
        \Else
           \State $\transact \gets \text{true}$
           \State $\aminforcenew \gets \aminforcee$
      \EndIf
\EndIf
\If{$\aminforcenew > \acommm\colon$} $\Call{DiscardPreviousMessages}{}()$
\EndIf
\State $\acommm \gets \aminforcenew$
\State $\Call{SendNewLimitToSucceedingVeh}{\acommm}$
\LineComment{\textbf{Process incoming safety request:} Case A of vehicle $i+1$}
\State safe$\gets \text{true}$
\vtab \If{$\alead < \aleadstored\colon$}
\vtab    \State safe~$\gets \verify(-\infty, \{i+1\})$ using $\alead$ for $\aleadstored$
\EndIf
\If{safe$\colon$} $\aleadstored \gets \alead$
\vtab \Else: $\transact \gets \text{true}$
\EndIf
\State $\Call{SendConfirmationToPrecedingVeh}{\aleadstored}$
\LineComment{\textbf{Process received safety confirmation:}}
\vtab \If{$\aconff \leq \aminforcee\colon$}
\vtab \State $\aminforcee \gets \max(\{\aconff, \acommm\})$
\EndIf
%
\State \Return $\getatrans(\transact)$
\end{algorithmic}
\end{algorithm}
    \vspace{-0.3cm}
\end{figure}
Let us now examine crucial steps of Alg.~\ref{alg:update-a-min}.

\paragraph{Apply consensus scheme}
Apart from sending a safety request in case A,
vehicle $i$ also needs to send its currently used braking limit $\aminforcee$ to the succeeding vehicle in case B
(ll. 9, 11-12).
If the braking limit sent to vehicle $i-1$ is larger than in the previous planning step,
vehicle $i$ discards all messages sent so far (l. 10).
The following proposition, which is used in the proof of Lemma~\ref{lemma:a-min-inv}, is thus fulfilled:
\begin{proposition}\label{prop:time-reset} 
If at times $t' < t$, vehicle $i$ sent the braking limits $\varat{\acommm}{t', i} < \varat{\acommm}{t, i}$, respectively,
in line 12, a confirmation for the braking limit $\varat{\acommm}{t', i}$ received at time $t$ is discarded.
\end{proposition}

\paragraph{Process incoming safety request}
If $\alead < \aleadstored$ (l. 15), vehicle $i$ knows that vehicle $i+1$ requests a confirmation for $\alead$,
and re-verifies its safety.
If safety is provided, vehicle $i$ updates the stored braking limit $\aleadstored$ for vehicle $i+1$ (l. 17).
It always sends a confirmation for the currently stored braking limit $\aleadstored$ (l. 19), which is either the currently
requested one, or was requested previously.
By that, we account for lost confirmations in case communication fails.

\paragraph{Process received safety confirmation}
As a confirmation by the succeeding vehicle is only required if $\aminforcenew < \aminforcee$
(cf. case A in Fig.~\ref{fig:details-consensus-algo}), a confirmation for $\aconff > \aminforcee$ is obviously outdated
and thus ignored (l. 21).
The adopted braking limit is the maximum of the confirmed braking limit and the
currently communicated one (l. 22).
By that, the last braking limit provided by the consensus entity is not undercut.

\paragraph{Increasing distance to front}
A vehicle executing a safety verification unsuccessfully must start to increase the distance to the
preceding vehicle.
This is indicated by the variable $\transact$,
which is always disabled at the beginning of Alg.~\ref{alg:update-a-min} (l. 1), and
possibly enabled later (ll. 8 and 18).
If $\transact$ is true, the function $\getatrans$ (l. 23) returns a successively decreasing
acceleration starting from the currently applied one $a(t_{k-1})$, which is used as an upper
bound for the input acceleration (cf. l. 5 in Alg.~\ref{alg:overall-algo}).
If $\transact$ is false, $\getatrans$ returns $\infty$.
%
%
%
\subsection{Safety Proof}\label{subsec:consensus-proof}
This section proves that Alg.~\ref{alg:update-a-min} ensures safe braking limit changes.
We explicitly assume that the vehicles of the platoon execute Alg.~\ref{alg:update-a-min} concurrently, and
we take communication failures and delays into account by not assuming that messages always and immediately
arrive or arrive in the order sent.
By providing each message with a timestamp, the vehicles can filter out obsolete messages as noted in the following Remark:
\begin{remark}\label{remark:order-preserving}
A vehicle always discards a message received by another vehicle if it received a more recent message by
that vehicle before.
\end{remark}
%
Note that an unexpected termination of an execution of Alg.~\ref{alg:update-a-min} at any point does not impede safety.
To achieve this, we never assume in the proofs below that the execution of one line in Alg.~\ref{alg:update-a-min}
implies the execution of a subsequent one.
Throughout this section, we write $\varat{\square}{t,i}$ to refer to the value of variable $\square$
at time $t$ during the execution of Alg.~\ref{alg:update-a-min} by vehicle $i$.
We make use of the following lemmas:
\begin{lemma}\label{lemma:braking-limit-inv} 
It always holds that $\acommm \leq \aminforcee$.

\proof{
    We only change $\acommm$ in line 11, where the new value is $\aminforcenew$.
    In case the \textit{if}-block in line 4 was not executed, it holds that $\aminforcenew < \aminforcee$.
    In case it was executed, $\aminforcenew=\aminforcee$ holds afterwards. $\square$
}
\end{lemma}
\begin{lemma}\label{lemma:ego-braking-limit-changes} 
$\aminforcee$ is never decreased in line 6 and never increased in line 22.

\proof {
The former case is implied by the condition in line 4.
The latter case follows from $\aconff \leq \aminforcee$ (l. 21) and Lemma~\ref{lemma:braking-limit-inv}. $\square$
}
\end{lemma}
%

\begin{lemma}\label{lemma:a-min-inv}
At any time $t$, the braking limit that vehicle $i-1$ assumes for its coupled preceding vehicle $i$ is an underestimation:
\begin{equation}\label{eq:inv-a-min}
\varat{\aleadstored}{t, i-1} \leq \varat{\aminforcee}{t, i} \text{.}
\end{equation}

\proof
{
    see Appendix.
}
\end{lemma}

\begin{lemma}\label{lemma:ego-braking-limit}
If a vehicle is safe w.r.t. each preceding vehicle and changes $\aminforcee$,
it is also safe afterwards.

\proof {
    Line 5 implies safety for the change in line 6, and
    Lemma~\ref{lemma:ego-braking-limit-changes} for the one in line 22.
    $\square$
}
\end{lemma}
\begin{lemma}\label{lemma:leader-braking-limit}
If a vehicle is safe w.r.t. its preceding vehicle applying $\aleadstored$, and the
vehicle changes $\aleadstored$, it is also safe if the preceding vehicle uses the new value of $\aleadstored$.

\proof {
    A vehicle can change $\aleadstored$ only in line 17, where for the new value $\alead$ it either holds
    that $\alead \geq \aleadstored$, or safety verification was done in line 16. 
    $\square$
}
\end{lemma}
We can now prove the safety of the protocol.
\begin{theorem}[Braking Limit Changes are Safe]\label{theorem:safe-changes}
If vehicle $i$ changes $\aminforcee$ in Alg.~\ref{alg:update-a-min}, safety of
1) vehicle $i$ w.r.t. each vehicle $j > i$, and safety of 2) vehicle $i-1$ w.r.t. vehicle $i$ is maintained.

\proof {
Lemma~\ref{lemma:ego-braking-limit} proves part 1).
Part 2) holds as the braking limit that vehicle $i-1$ assumes about vehicle $i$ is always underapproximate according to
Lemma~\ref{lemma:a-min-inv}, and a change of $\aleadstored$ always preserves safety with
    Lemma~\ref{lemma:leader-braking-limit}.
}
\end{theorem}

%
\section{EVALUATION}\label{sec:eva}
We evaluate our concept on various scenarios using the CommonRoad platform~\cite{althoff_commonroad_2017}.
We executed all simulations on a machine with an AMD Ryzen 9 5900HX processor with 4.6GHz and 64 GB of DDR4 3.200 MHz
memory, and we implemented the safety protocol in Python.
Tab.~\ref{tab:all-params} shows the values of common parameters.
We use a PD controller as nominal CACC inspired by~\cite{baldi_establishing_2021} that maintains
a headway $h=\SI{0.3}{\s}$ to the preceding vehicle or controls a certain velocity.
Furthermore, we use the consensus scheme proposed in~\cite{liu_resilience_nodate}, extended by a simple reset
mechanism if a vehicle leaves the platoon to enable a decreasing convergence target for the braking limit.
\begin{table}
    \vspace{0.2cm}
\caption{Common Parameters}
\label{tab:all-params}
\centering
    \begin{tabular}{l l l l}
    \toprule
    Parameter & Value & Parameter & Value\\
    \midrule
    $[\underline{\rho},\overline{\rho}]$ & $[1.1, 1.3]\si{\kilogram\per\cubic\metre}$ & $\tplanning$ & $\SI{0.1}{\second}$ \\
    \rule{0pt}{0.3cm}$[\underline{v}{}_{\text{wind}},\overline{v}_{\text{wind}}]$ & $[1.4, 4.2]\si{\metre\per\second}$ & $a_{\text{tol}}$ & $\SI{0.05}{\metre\per\square\second}$ \\
    $[\underline{\alpha},\overline{\alpha}]$ & $[-0.06, 0.06]\rad$ & $a_{\text{cut-in}}$ & $\SI{1}{\metre\per\square\second}$ \\
    $[\wmingeneral,\wmaxgeneral]$ & $[-0.1, 0.1]~\si{\metre\per\square\second}$~\cite{schurmann_ensuring_2017} & $t_C$ & $\SI{4}{\second}$ \\
    $\sensorrange$ & $\SI{200}{\metre}$  & & \\
    \bottomrule
    \end{tabular}
    \vspace{0.1cm}
\end{table}
For the simulations, we generate measurement intervals and disturbances with a Gaussian distribution within the
specified range, employing a 99\% confidence interval and truncating values outside this range.
We assume a measurement uncertainty range of $\pm\vectorhoriz{\SI{0.2}{\metre} & \SI{0.05}{\metre\per\second}}$
for the state of the ego vehicle, and $\pm\vectorhoriz{\SI{0.1}{\metre} & \SI{0.05}{\metre\per\second}}$
for the relative states of the preceding vehicles~\cite{engels_automotive_2021}.
Tab.~\ref{tab:vehicles-params} shows the vehicle parametrizations used in the following scenarios.
The runtime of a single planning step was consistently below $\SI{80}{\milli\second}$,
highlighting the real-time capability of the approach.
For each scenario, we show the occupancies relative to the first vehicle,
and the position induced by the safe distances to the direct predecessor (dashed).
We also plot the effective accelerations, already including disturbances;
we use dotted lines to indicate that the fallback controller was active.
\begin{table}[t]
    \vspace{0.2cm}
\caption{Vehicle parameters~\cite{pischinger_vieweg_2021} }
\label{tab:vehicles-params}
\centering
\begin{tabular}{l l l l l l l }
\toprule
Parameter & Worst Case & $p_0$ & $p_1$ & $p_2$ & $p_3$ & $p_4$ \\
\midrule
     $\aminforcee [\si{\metre\per\square\second}]$ & $-12$     & $-5$      & $-6$      & $-10$     & $-5.5$    & $-9$    \\
     \rule{0pt}{0.3cm}$\amaxforcee [\si{\metre\per\square\second}]$ & ---       & $1$       & $1.5$     & $4$       & $1$       & $3.5$   \\
     \rule{0pt}{0.3cm}$\vmaxx [\si{\metre\per\second}]$        & ---       & $25$      & $25$      & $60$      & $25$      & $50$    \\
     $m [\si{\tonne}]$                  & $0.4$     & $20$      & $15$      & $2.5$     & $20$      & $2$     \\
     $c$                 & $2$       & $0.7$     & $0.5$     & $0.25$    & $0.6$    & $0.35$     \\
     $A [\si{\square\metre}]$         & $12.5$    & $7$       & $8$       & $1.7$     & $6$      & $2.4$   \\
     $l [\si{\metre}]$                 & ---       & $16$      & $14$      & $4.9$     & $16$      & $4.2$     \\
\bottomrule
\end{tabular}
\end{table}

\subsection{Scenario 1: Fallback Controller Evaluation}
We consider a scenario with a platoon of two trucks parametrized by $p_0$ and $p_1$
(cf. Tab.~\ref{tab:vehicles-params}), where no consensus technique is used.
The platoon is slowly approaching a non-platoon vehicle, which performs a full brake starting at $t=\SI{30}{\second}$.
The results are shown in Fig.~\ref{fig:scenario1}.
Vehicles $1$ and $2$ always keep a safe distance, and no collision occurs when vehicle $3$ brakes fully.
The nominal CACC frequently fails to compute a safe input for vehicle $1$, so the fallback controller is engaged.
The inputs computed by the latter are usually above $\SI{-1}{\metre\per\square\second}$,
confirming that the fallback controller prevents unnecessarily harsh safety interventions.
\begin{figure}
    \resizebox{\columnwidth}{!}{\input{fail_safe_demo_2_all.pgf}}
    \vspace{-7mm}
  \caption{(a) Occupancies. The safe distance drops after coupling at the very beginning.
           (b) Effective accelerations and $\aminofi$ (dashed) for each platoon vehicle $i$, and road incline
           angle $\alpha$ (red) from the perspective of $2$.}
  \label{fig:scenario1}
    \vspace{1mm}
\end{figure}

\subsection{Scenario 2: Consensus Techniques}
\begin{figure}
    \vspace{0.1cm}
\begin{tabular}{m{0.45cm} c }
    {\small (a)} &
    \begin{minipage}{0.86\columnwidth}
      \includegraphics[width=\columnwidth]{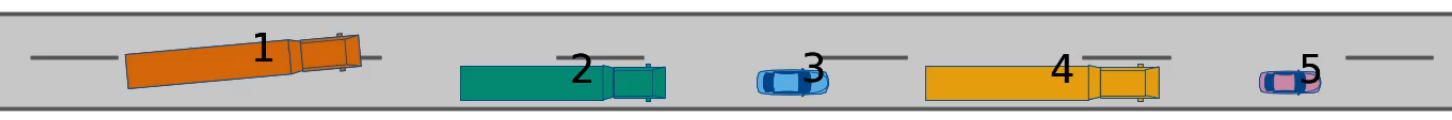}
    \end{minipage}
    \\
\end{tabular}
\resizebox{\columnwidth}{!}{\input{consensus_3_all.pgf}}
\vspace{-7mm}
\caption{(a) Illustration of CommonRoad scenario at $t=\SI{26}{\second}$. (b) Occupancies
(c) Velocities
(d) Effective accelerations and $\aminforce{i}$ (dashed).
      After vehicle $1$ leaves the platoon, the consensus braking limit decreases.}
\label{fig:scenario2}
\vspace{1mm}
\end{figure}

We consider a scenario with a platoon of three trucks and two cars heterogeneously parametrized by $p_0$ to $p_4$
(cf. Tab.~\ref{tab:vehicles-params}).
The scenario is executed with the consensus scheme.
The results are shown in Fig.~\ref{fig:scenario2}.
A much denser spacing is achieved after the consensus scheme converges at $t\approx\SI{5}{\second}$.
Vehicle $1$, which has the weakest braking capability, leaves the platoon at $t=\SI{29}{\second}$ by a lane
change, resulting in a reset of the consensus and a stronger consensus braking limit formed afterward.
The full brake performed by the leader at $t=\SI{80}{\second}$ does not end in a collision.
The safe distances are always kept despite the dense spacing.

\section{CONCLUSION}\label{sec:concl}
We propose a provably correct safety protocol for cooperative platooning that can be applied to any existing CACC.
Considering changing road inclines,
the safety of each nominal input is verified online against every possible acceleration behavior of the preceding
vehicles, which is enabled by dynamics that is monotone in the position domain.
If the nominal input is identified as unsafe, a fallback controller prevents harsh interventions.
Impending collisions caused by vehicles cutting in are communicated backward to allow
the succeeding vehicles to react proactively.
For consensus techniques establishing a common braking limit among the platoon vehicles,
we additionally propose a protocol allowing the vehicles to safely change their braking limits.
We confirm the benefit of our approach in our experiments.

\section*{ACKNOWLEDGMENT}
The authors thank Adrian Kulmburg for his help with the proof of Theorem~\ref{prop:monotonicity}.
Furthermore, the authors gratefully acknowledge the financial support by the European Commission Project justITSELF
under grant number 817629.

\bibliographystyle{IEEEtran}

\bibliography{IEEEabrv,bibl,further}

\appendix
\section{Appendix}

\subsection{Details of Computing Bounds on Reachable Positions}
When computing the position bounds $\lowerpos{j}(a_d(\cdot),t)$ and $\upperpos{i}(a_d(\cdot),t)$, we need to
evaluate $\aminn(\alpha,v)$ and $\amaxx(\alpha,v)$ (cf.~\eqref{eq:overalllimits}) under- and overapproximatively, respectively.
With the global acceleration limits
\begin{equation*}
    \begin{aligned}
        & \aminglobal=\aminforcee+\incline(\overline{\alpha})+\adrag(\vmaxx)~\text{and} \\
        & \amaxglobal=\amaxforcee+\incline(\underline{\alpha})\text{,}
    \end{aligned}
\end{equation*}
we overapproximate the reachable position, velocity and incline intervals during the time $\tstep$ by
\begin{equation*}
    \begin{aligned}
        & \mathcal{I}_s=[s,s+v\tstep+0.5\amaxglobal\tstep^2]\text{,} \\
        & \mathcal{I}_v=[v+\tstep\aminglobal,v+\tstep\amaxglobal],~\text{and} \\
        & \mathcal{I}_{\alpha}=[\min_{s'\in \mathcal{I}_s}\alpha(s'),\max_{s'\in \mathcal{I}_s}\alpha(s')]
        \text{.}
    \end{aligned}
\end{equation*}
We evaluate $\aminn(\alpha,v)$ and $\amaxx(\alpha,v)$ on the right limits of
$\mathcal{I}_{\alpha}$ and $\mathcal{I}_v$ for computing $\lowerpos{j}(a_d(\cdot),t)$,
and on the left limits for computing $\upperpos{i}(a_d(\cdot),t)$.

\subsection{Proof of Lemma~\ref{lemma:a-min-inv}}
Initially,~\eqref{eq:inv-a-min} holds due to the worst-case assumption that vehicle $i-1$ makes about
vehicle $i$.
At time $t$,~\eqref{eq:inv-a-min} can only change if
vehicle $i$ has changed $\aminforcee$ (l. 6 or 22), or
vehicle $i-1$ has changed $\aleadstored$
(l. 17 in Alg.~\ref{alg:update-a-min}).
Let us examine these three cases:

\textbf{Vehicle $i$ has changed $\aminforcee$ in line 6}:
This does not change~\eqref{eq:inv-a-min} according to Lemma~\ref{lemma:ego-braking-limit-changes}.
\begin{figure}[th]
    \begin{tabular}{m{0.2cm} c } 
    {\small (a)} &
    \begin{minipage}{0.79\columnwidth}
        \hspace{-0.4cm}
        \resizebox{\columnwidth}{!}{

\tikzset{
    arr/.style   = {-{Latex[length=3mm]}},
    comm/.style = {arr, dash pattern=on 3pt off 3pt}, 
    dotted/.style = {dash pattern=on 6pt off 6pt}, 
    arrrev/.style   = {{Latex[length=3mm]}-},
}

\newcommand{\usedfs}{\nfs}
\newcommand{\usedsmallfs}{\sfs}
\newcommand{\tickoffsetwest}{6pt}
\newcommand{\tickoffseteast}{-8pt}
\newcommand{\textonlineoffset}{13pt}

\begin{tikzpicture}[
    node distance=1cm,
    every node/.style={inner sep=0,outer sep=0},
    label/.style={font=\sffamily},
]
    \coordinate (sb) at (0,0);
    \coordinate (st) at (0,5);
    \coordinate (pb) at (10,0);
    \coordinate (pt) at (10,5);

    \draw[arrrev] (sb) -- (st) node[pos=1.1]{\usedfs Vehicle $i-1$};
    \draw[arrrev] (pb) -- (pt) node[pos=1.1]{\usedfs Vehicle $i$};

    \draw ($(pb)!0.85!(pt)+(-2pt,0)$) node(t0) {} -- ($(pb)!0.85!(pt)+(2pt,0)$)
    node[anchor=west, xshift=\tickoffsetwest] {\usedfs $t'$}
    node[anchor=west, xshift=\tickoffsetwest+18pt] {\circled{\usedfs \textbf{3}}};; 

    \draw ($(sb)!0.65!(st)+(-2pt,0)$) -- ($(sb)!0.65!(st)+(2pt,0)$)
    node[anchor=east, xshift=\tickoffseteast] {\usedfs $t_a$} node(t1) {}
    node[anchor=east, xshift=\tickoffseteast-20pt] {\circled{\usedfs \textbf{2}}};

    \draw ($(sb)!0.25!(st)+(-2pt,0)$) -- ($(sb)!0.25!(st)+(2pt,0)$) node[anchor=east, xshift=\tickoffseteast] {\usedfs $t_b$} node(t3) {}
    node[anchor=east, xshift=\tickoffseteast-20pt] {\circled{\usedfs \textbf{4}}};;

    \draw ($(pb)!0.4!(pt)+(-2pt,0)$) node(t) {} -- ($(pb)!0.4!(pt)+(2pt,0)$) node[anchor=west, xshift=\tickoffsetwest] {\usedfs $t_c$} node(t2) {}
    node[anchor=west, xshift=\tickoffsetwest+18pt] {\circled{\usedfs \textbf{5}}};

    \draw ($(pb)!0.15!(pt)+(-2pt,0)$) node(t) {} -- ($(pb)!0.15!(pt)+(2pt,0)$) node[anchor=west, xshift=\tickoffsetwest] {\usedfs $t$}
    node[anchor=west, xshift=\tickoffsetwest+18pt] {\circled{\usedfs \textbf{1}}};

    \draw[comm] (t0) -- node[midway,sloped, anchor=center, yshift=\textonlineoffset]{\usedfs $\varat{\acommm}{t', i}$} (t1);
    \draw[comm] (t1) -- node[midway,sloped, anchor=center, yshift=\textonlineoffset]{\usedfs $\varat{\aconff}{t, i}$} (t);
    \draw[comm] (t2) -- node[pos=0.8,sloped, anchor=center, yshift=\textonlineoffset]{\usedfs $\varat{\acommm}{t_c, i}$} (t3);
    \draw[dotted] ($(sb)!0.15!(st)+(-2pt,0)$) -- (t);
\end{tikzpicture}}
    \end{minipage}
    \\
    & \\
    {\small (b)} &
    \begin{minipage}{0.79\columnwidth}
        \hspace{-0.4cm}
        \resizebox{\columnwidth}{!}{

\tikzset{
    arr/.style   = {-{Latex[length=3mm]}},
    arrrev/.style   = {{Latex[length=3mm]}-},
    comm/.style = {arr, dash pattern=on 3pt off 3pt}, 
    dotted/.style = {dash pattern=on 6pt off 6pt}, 
}

\newcommand{\usedfs}{\nfs}
\newcommand{\usedsmallfs}{\sfs}
\newcommand{\tickoffsetwest}{6pt}
\newcommand{\tickoffseteast}{-8pt}
\newcommand{\textonlineoffset}{13pt}

\begin{tikzpicture}[
    node distance=1cm,
    every node/.style={inner sep=0,outer sep=0},
    label/.style={font=\sffamily},
]
    \coordinate (sb) at (0,0);
    \coordinate (st) at (0,5);
    \coordinate (pb) at (10,0);
    \coordinate (pt) at (10,5);

    \draw[arrrev] (sb) -- (st) node[pos=1.1]{\usedfs Vehicle $i-1$};
    \draw[arrrev] (pb) -- (pt) node[pos=1.1]{\usedfs Vehicle $i$};


    \draw ($(pb)!0.85!(pt)+(-2pt,0)$) node(t0) {} -- ($(pb)!0.85!(pt)+(2pt,0)$) node[anchor=west, xshift=\tickoffsetwest] {\usedfs $t_b'$}
    node[anchor=west, xshift=\tickoffsetwest+18pt] {\circled{\usedfs \textbf{5}}};

    \draw ($(sb)!0.65!(st)+(-2pt,0)$) -- ($(sb)!0.65!(st)+(2pt,0)$) node(t1) {} node[anchor=east, xshift=\tickoffseteast] {\usedfs $t_c$}
    node[anchor=east, xshift=\tickoffseteast-20pt] {\circled{\usedfs \textbf{4}}};

    \draw ($(pb)!0.45!(pt)+(-2pt,0)$) node(t) {} -- ($(pb)!0.45!(pt)+(2pt,0)$) node[anchor=west, xshift=\tickoffsetwest] {\usedfs $t_a$} node(tp) {}
    node[anchor=west, xshift=\tickoffsetwest+18pt] {\circled{\usedfs \textbf{2}}};

    \draw ($(sb)!0.15!(st)+(-2pt,0)$) -- ($(sb)!0.15!(st)+(2pt,0)$) node(t) {} node[anchor=east, xshift=\tickoffseteast] {\usedfs $t$}
    node[anchor=east, xshift=\tickoffseteast-20pt] {\circled{\usedfs \textbf{1}}};

    \draw ($(pb)!0.25!(pt)+(-2pt,0)$) node(tpp) {} -- ($(pb)!0.25!(pt)+(2pt,0)$) node[anchor=west, xshift=\tickoffsetwest] {\usedfs $t_b$}
    node[anchor=west, xshift=\tickoffsetwest+18pt] {\circled{\usedfs \textbf{3}}};

    \draw[comm] (t0) -- node[midway,sloped, anchor=center, yshift=\textonlineoffset]{\usedfs $\varat{\acommm}{t_b', i}$} (t1);
    \draw[comm] (t1) -- node[midway,sloped, anchor=center, yshift=\textonlineoffset]{\usedfs $\varat{\aconff}{t_b, i}$} (tpp);
    \draw[comm] (tp) -- node[pos=0.75,sloped, anchor=center, yshift=\textonlineoffset]{\usedfs $\varat{\acommm}{t_a, i}$} (t);
    \draw[dotted] (t) -- ($(pb)!0.15!(pt)+(2pt,0)$);
\end{tikzpicture}}
    \end{minipage}
    \end{tabular}
    \caption{Communication processes illustrating the proofs of Lemma~\ref{lemma:a-min-inv}
    (a) Case 2, and (b) Case 3.}
    \label{fig:proof-seq}
\end{figure}

\textbf{Vehicle $i$ has changed $\aminforcee$ in line 22} (cf. \circled{\textbf{1}} in Fig.~\ref{fig:proof-seq}~(a)):
Let $t_a \leq t$ be the time at which vehicle $i-1$ sent the confirmation for $\varat{\aconff}{t, i}$
(cf. \circled{\textbf{2}} in Fig.~\ref{fig:proof-seq}~(a)),
and $t'\leq t_a$ the last time that vehicle $i$ communicated $\varat{\acommm}{t', i}=\varat{\aconff}{t, i}$
(cf. \circled{\textbf{3}} in Fig.~\ref{fig:proof-seq}~(a)).
Then, vehicle $i-1$ stores $\varat{\aconff}{t, i}$ in its variable $\aleadstored$.
As $\varat{\aconff}{t, i} \leq \varat{\aminforcee}{t, i}$ (cf. l. 22),
~\eqref{eq:inv-a-min} holds at time $t$,
unless vehicle $i-1$ increased $\aleadstored$ in line 17 at some time $t_b \in [t_a, t]$,
which we assume for the sake of contradiction
(cf. \circled{\textbf{4}} in Fig.~\ref{fig:proof-seq}~(a)).
Then, vehicle $i$ must have sent a braking limit at some time $t_c \in [t', t_b]$
with $\varat{\acommm}{t_c, i} > \varat{\acommm}{t', i}$
(cf. \circled{\textbf{5}} in Fig.~\ref{fig:proof-seq}~(a)).
However, Prop.~\ref{prop:time-reset} gives that vehicle $i$ would not have accepted the confirmation
received at time $t$ then, thus our assumption was wrong.

\textbf{Vehicle $i-1$ has changed $\aleadstored$ in line 17}
(cf. \circled{\textbf{1}} in Fig.~\ref{fig:proof-seq}~(b)):
Then vehicle $i$ sent $\varat{\acommm}{t_a, i}=\varat{\aleadstored}{t, i-1}$ in line 12 at some time $t_a \leq t$
(cf. \circled{\textbf{2}} in Fig.~\ref{fig:proof-seq}~(b)).
As $\varat{\acommm}{t_a, i} \leq \varat{\aminforcee}{t_a, i}$ due to Lemma~\ref{lemma:braking-limit-inv},
~\eqref{eq:inv-a-min} holds at time $t$, unless vehicle $i$ decreased $\aminforcee$ to a value below
$\varat{\acommm}{t_a, i}$ at some time $t_b \in [t_a, t]$ in line 22,
which we assume for the sake of contradiction
(cf.~\circled{\textbf{3}} in Fig.~\ref{fig:proof-seq}~(b)),
i.e., $\varat{\aminforcee}{t_b, i}<\varat{\acommm}{t_a, i}$.
Let $t_c$ the time that vehicle $i-1$ sent the confirmation for $\varat{\aconff}{t_b, i} \leq \varat{\aminforcee}{t_b, i}$
(cf.~\circled{\textbf{4}} in Fig.~\ref{fig:proof-seq}~(b)), and
$t_b'\leq t_c$ the last time that vehicle $i$ sent $\varat{\acommm}{t_b', i}=\varat{\aconff}{t_b, i}$ to vehicle $i-1$
(cf. \circled{\textbf{5}} in Fig.~\ref{fig:proof-seq}~(b)).
It holds that $t_b'<t_a$ with Remark~\ref{remark:order-preserving}.
Now we have that vehicle $i$ sent $\varat{\acommm}{t_b', i} < \varat{\acommm}{t_a, i}$ at times $t_b' < t_a$, respectively.
However, Prop.~\ref{prop:time-reset} gives that vehicle $i$ would not have accepted the confirmation
received at time $t_b$ then, thus our assumption was wrong. $\square$

\end{document}